# Many-body effects observed in the positron annihilation experiment

G. Kontrym-Sznajd[1] and H. Sormann[2],

[1]Institute of Low Temperature and Structure Research, Polish Academy of Sciences, P. O. Box 1410, 50-950 Wroclaw 2, Poland, email: gsznajd@int.pan.wroc.pl

[2]Institut für Theoretische Physik- Computational Physics, Technische Universität Graz, Petersgasse 16, A-8010 Graz, Austria, email: sormann@itp.tu-graz.ac.at

This paper is devoted to study many-body effects in the positron annihilation experiment, both electron-positron (*e-p*) and electron-electron (*e-e*) correlations. Various theories of the *e-p* interaction in real solids were used to verify them by comparing theoretical and experimental *e-p* momentum densities in Cu and Y. We show that the lattice potential has an essential influence on the *e-p* correlation effects, i.e. their proper description must be done via periodic lattice potential as e.g. in the Bloch Modified Ladder (*BML*) theory. Moreover, it is not true that that the dynamic parts of the direct *e-p* and *e-e* interactions cancel each other because *e-e* correlations are observed not only in the Compton scattering but also in the positron annihilation experiments.

Key words: positron annihilation, Compton scattering, momentum densities, many-body effects

## 1. Introduction

The first question considered in the paper is connected with the electron-positron (*e-p*) interaction in real metals. It is shown that the Bloch-modified ladder (*BML*) theory [1], in contrast to all other approaches, is able to describe (at least qualitatively) experimental *e-p* momentum densities for both simple and transition metals. Such a finding is important because all these theories, except *BML*, ignore the influence of the lattice potential on the *e-p* interaction, concerning both intraband and interband transitions.

The next subject of the paper are the electron-electron (*e-e*) correlations which should, in principle, be also observed in the positron annihilation experiments. However, Carbotte and Kahana [2] showed that (at least for positrons in jellium) *the dynamical parts* of the *e-p* and *e-e* correlation cancel each other. Consequently, the remaining many-body effects come only from the *static* part of these interactions. Because the static *e-e* correlations are (at least approximately) included into the band structure calculations, almost all positron annihilation theories consider only the static part of the *e-p* correlations, based on the result of Carbotte and Kahana [2]: an annihilating *e-p* pair is, seen from outside, a neutral quantity with a strongly reduced coupling to its environment ([3] and Ref. therein). The resulting enhancement factor is strongly momentum-dependent and leads to a monotonously increasing *e-p* momentum density below the Fermi momentum $p_F$, an effect which we call *Kahana-like* enhancement. Beyond that, there exists only one electron gas theory by Arponen and Pajanne [4] where the *e-e* interaction, on the level of the well-known random phase approximation (*RPA*), is described by non-interacting Sawada bosons, and each boson-boson interaction goes beyond the *RPA*. Contrary to the result of Ref. [2], Arponen and Pajanne observed a significant tail of the *e-p* momentum density beyond $p_F$ due to dynamical *e-e* and *e-p* correlations. However, according to the results in Ref. [4], the values of the enhancement factor (*EF*) on the Fermi surface (*FS*) increase with the increasing density of the electron gas. Since such a behaviour of the EF strongly contradicts the experiments, all theories of the *e-p* annihilation [3] are based on the results of Kahana and Carbotte [2].

According to our knowledge, the existence of many-body tails in positron annihilation data has been observed for the first time by Manuel *et al.* [5] for Sn-β and even for such *jellium-like* metals as Li and Al. Next, Ohata *et al.* [6] compared a high-resolution Compton profile (CP) with one–dimensional (1D) angular correlation of annihilation radiation (ACAR) spectrum along [111] direction in Al. They observe a *Kahana-like* enhancement near the FS and a weaker tail of densities for $p > p_F$ in the case of the 1D ACAR data compared to the CPs, as a consequence of the partial cancellation of *e-e* and *e-p* correlations. Here we would like to point out that the BML theory, applied to Al [1(a)], gives the following results. Whereas the enhancement factor for momenta $p < p_F$ is similar to the *Kahana-like* enhancement, for $p > p_F$ the contribution of Umklapp components is significantly diminished by the *e-p* interaction. Moreover, it reduces (in comparison with IPM) the core contribution as well as the enhancement factor for core electrons decreases for higher momenta. So, a weaker tail for $p > p_F$ observed in Al [4], could be connected with these *e-p* correlation effects (not with weaker *e-e* correlations as interpreted in [6]). A simultaneous analysis of both reconstructed densities and 1D profiles for Compton scattering and 2D ACAR experiment in Y, allowed us to state that in this material *e-e* correlations in the ACAR data are exactly the same as in the Compton scattering experiment [7], an effect which has been recently observed also in LaB$_6$ [8] and in Mg [9].

## 2. Applied theories.

In the angular correlation of annihilation radiation (*ACAR*) or Compton scattering experiments one measures integrals of the electron-positron or electron momentum densities in the extended *p* space, respectively

$$\rho(\mathbf{p}) = \sum_j n_{\mathbf{k}j} \left| \int_{-\infty}^{\infty} e^{-i\mathbf{p}\cdot\mathbf{r}} \psi_{\mathbf{k}j}^{e-p}(\mathbf{r},\mathbf{r}) d\mathbf{r} \right|^2, \qquad (1)$$

where $n_{kj}$ is the occupation number (0 or 1) of the electron Bloch state *kj* and $\psi_{\mathbf{k}j}^{e-p}(\mathbf{r},\mathbf{r})$ is the pair wave function of an electron and a thermalized positron.

We used five models for the wave function of an *e-p* pair moving within a lattice-periodical crystal potential, where for the numerical evaluation of the electron and positron wave functions, the augmented plane-wave (APW) method has been applied (details in [10]):

1. $\psi_{kj}^{e-p}(r,r) = \psi_{kj}(r)$ - the electron momentum density (*EMD*).

2. $\psi_{kj}^{e-p}(r,r) = \psi_{kj}(r)\psi_+(r)$ - the independent particle model (*IPM*).

3. $\psi_{kj}^{e-p}(r,r) = \sqrt{g(r,kj)}\psi_{kj}(r)\psi_+(r)$ - the most popular approach, where a local *e-p* correlation function *g* is inserted into IPM formula. In the present paper, for *g* the following cases were considered:

3.1. the local-density approximation (*LDA*), as proposed by Daniuk *et al.* [11] where $g(r; jk) = \varepsilon_{hom}(r_s(r); \chi_{jk})$ with $r_s(r)$ as the *local density* parameter. The enhancement factors $\varepsilon_{hom}$ are results of an *e-p* enhancement theory for the electron gas [12]. Here the correlation function is state-dependent " (we call it state-dependent LDA) where $\chi_{nk} = [(E_{jk}-E_0)/(E_F-$ $E_0)]^{1/2}$ with $E_0$, $E_F$ and $E_{jk}$ as: the bottom energy of the electron conduction bands, Fermi energy and electron energy in the Bloch state $|jk>$, respectively.

3.2 An LDA-type theory which neglects the explicit momentum-dependence of correlation function $g$ (we call it state-independent LDA) [13]. As before, $\varepsilon_{hom}$ is taken for homogeneous electron gases (we applied the formula of Boroński and Nieminen [14]).

4. The so-called Bloch-modified ladder (BML) theory [1a], based on an earlier paper of Carbotte [1b] and Fujiwara [1c] where the *e-p* interaction is included via a lattice-periodic crystal potential.

Other theories, used for describing the *e-p* interaction in real metals (being similar to theories described in the point 3) are the following: the weighted density approximation (WDA) [15], the generalized gradient approximation (GGA) [16] and the theory proposed by Alatalo *et al.* and Barbiellini *et al.* [17] where the correlation function $g$ is substituted by the state dependent correlation factor $\gamma_{kj}$. However, this state dependence $k$ is not connected with either energy or momentum dependence as given in the function $g$. It follows from a state dependence of the ratio $\gamma_{kj} = \lambda_{kj}/\lambda_{kj}^{IPM}$, where $\lambda$ denotes the local an-hilation rates which could be calculated within state independent either LDA or GGA.

## 3. Results.

In this chapter we present theoretical *e-p* momentum densities $\rho(p)$ for yttrium and cooper, compared with densities reconstructed from both 2D ACAR spectra (for Y and Cu) and 1D high-resolution Compton profiles (for Y). 2D ACAR spectra represent line projections of *e-p* momentum densities while 1D Compton profiles (CPs) plane projections of electron densities, both densities are studied in the extended *p* space.

Five profiles for Y were measured with overall resolution about 0.15 atomic units of momentum (a.u.), using the 2D-ACAR spectrometer at the University of Texas at Arlington [18]. Next, *e-p* densities $\rho(p)$ were reconstructed by applying the Cormack's method [19]. *e-p* densities for Cu, reconstructed from six experimental 2D ACAR spectra, were taken from the Ref. [20]. Experimental electron momentum densities for Y were reconstructed from 12 high-resolution CPs, measured with overall resolution about 0.16 a.u. at the European Synchrotron Radiation Facility (ESRF), France – more details in [7].

In Fig. 1 we show results for Cu where the reconstructed densities shows a typical *Kahana-like* enhancement. This behaviour can be satisfactorily described by the state-dependent LDA and the BML theory, in contrast to the other theories mentioned in the previous chapter.

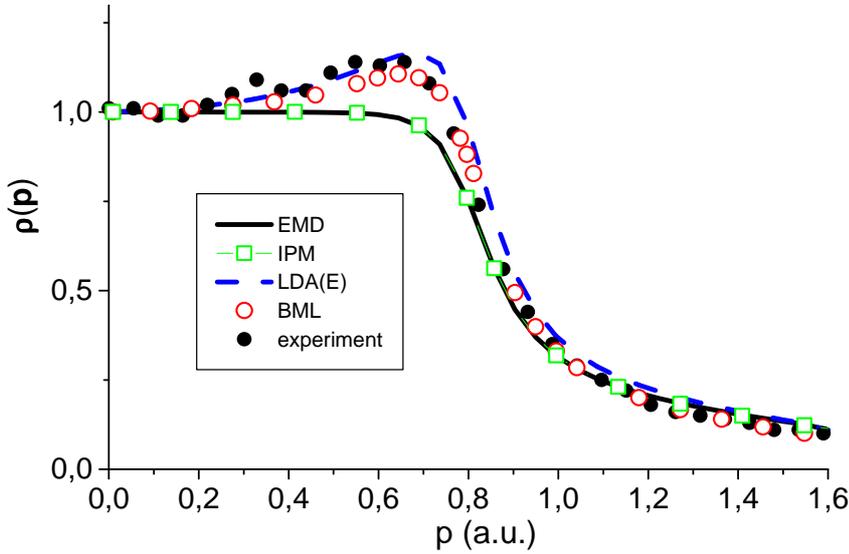

FIG. 1. Theoretical *e-p* momentum densities for copper along [111], compared with densities reconstructed from 2D ACAR data [20].

In Fig. 2 we present reconstructed *e-p* densities in Y along the direction ΓM on the basal ΓM K plane, compared with theoretical results. Presented theoretical results are not convoluted (not smeared by the experimental resolution) while reconstructed densities are after applying Max Entropy deconvolution procedure [18,21]. It is evident that the electron momentum density EMD is essentially different from the electron density "observed" by the positron, particularly for the high-momentum region. This behaviour reflects the well-known fact that high-momentum contributions to the momentum density are significantly reduced by the appearance of a positron.

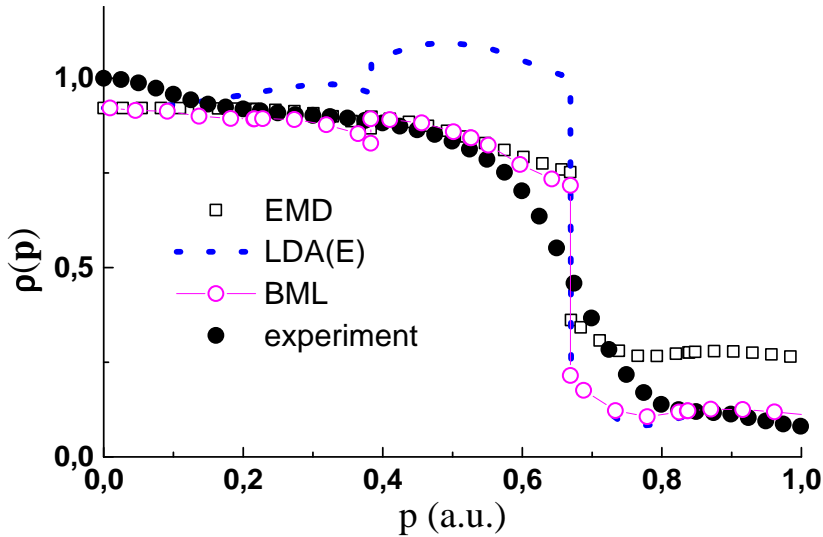

Fig. 2. Theoretical *e-p* momentum densities in Y along ΓM, compared with densities reconstructed from 2D ACAR experimental spectra.

For yttrium, the application of the IPM leads to a momentum profile which is very similar to the BML result (drawn by open circles), and both curves fit the reconstructed data rather well. The inclusion of *e-p* correlation effects shows the following behavior: the use of both a local and state-dependent correlation function (according to the proposal by Daniuk et al [11], see the dashed curve in Fig. 2), leads to a strongly increased momentum density. It is a typical *sp*-like enhancement, well-known for metals with marked nearly-free valence electrons like in alkalis, aluminium, or even copper [20]. However, for yttrium such a behaviour is in clear contrast to the (reconstructed) experimental densities. This is somewhat surprising, taking into account that the 5s+4d electrons in yttrium are far from the ions and should therefore be considered by the positrons as nearly-free particles (especially the electrons from the 1st and 2nd valence band where $E=E(\mathbf{k})$ is close to the parabolic function). The best agreement between theory and experiment can be obtained by including the *e-p* correlation effects according to a state-independent LDA with a local enhancement function, or by the application of the BML theory, particularly in those regions where contributions from the partially occupied 3rd and 4th band dominate. Still remaining differences between theory and experiment lead to the conclusion that contributions of Umklapp components of the electron wave function are stronger decreased by the positron than it follows from applied theories.

There is also no experimental indication for a *Kahana-like* enhancement [22,23] in a typical transition metal like Cr. Similarly as in Y, the momentum density decreases monotonously with increasing moment, and the density values yielded by a state-dependent LDA or by WDA [23] theory are too high, whereas other theories like state-dependent LDA or the BML approach lead to results which fit experiment rather well.

In Ref. 7, we recently published further results of Y, based on a simultaneous analysis of yttrium 2D-ACAR data and of high-resolution CPs which allowed us the observation of strong *e-e* correlation effects in the positron annihilation data. Theoretically, *e-e* correlations were taken into account following the work of Cardwell and Cooper [24] based on the proposal of Lam and Platzman [25], where the Lam-Platzman corrections have been calculated from the self-consistent APW electron charge density. Studying both a directional anisotropy of measured spectra and reconstructed densities, we obtained that in the case of Compton profiles there are strong *e-e* correlations which cannot be described by such isotropic *e-e* Lam-Platzman corrections – this effect is seen in Fig. 3b. However, it was even more surprising that we got **exactly the same** differences between theoretical and reconstructed experimental densities $\rho^{EMD}(\mathbf{p}) - \rho^{CP}(\mathbf{p})$ and $\rho^{IPM}(\mathbf{p}) - \rho^{ACAR}(\mathbf{p})$ (in Y $\rho^{IPM}(\mathbf{p})$ is almost the same as $\rho^{BML}(\mathbf{p})$, i.e. *e-p* correlations does not change momentum dependence of densities). So, it is clear that also the *e-p* momentum density are strongly influenced by the *e-e* correlations, presented clearly in figure 4d. This finding was surprising because almost all theories devoted to this question [3] are based on the result of Carbotte and Kahana [2] where *e-p* pair is, seen from outside, a neutral quantity with a strongly reduced coupling to its environment. Consequently, typical correlation effects as smearing at the Fermi momentum and high-momentum tails of the momentum distribution should be significantly small than in pure electron systems. However, a detailed analysis of 2D ACAR spectra show that it is not true.

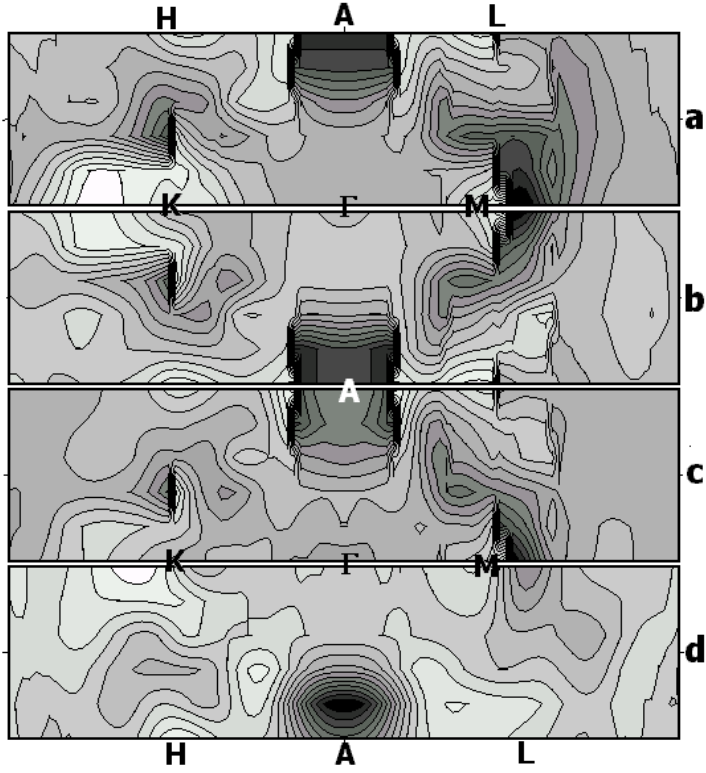

Fig. 3. Anisotropic part of electron momentum densities in Y along ΓM, ΓK and parallel directions (for momenta up to 1.37 a.u.) for:
(a) pure theoretical EMD densities; differences between theoretical and experimental densities:
(b) $\rho^{EMD}(\mathbf{p}) - \rho^{CP}(\mathbf{p})$;
(c) $\rho^{IPM}(\mathbf{p}) - \rho^{ACAR}(\mathbf{p})$;
(d) densities reconstructed from convoluted theoretical and experimental CPs.

## 4. Conclusions.

By applying various models of the *e-p* wave function, we found out that the *e-p* momentum density in simple metals as Al., with its typical Kahana-like enhancement, can be satisfactorily described by the use of state-dependent enhancement theories like state-dependent LDA or WDA. On the other hand, such theoretical approaches completely fail in the case of transition metals like Cr or Y where methods like the state-independent LDA succeed. Only the BML theory works reasonably well for both simple and transition metals. Our explanation for this behavior is as follows:

In the Kahana formalism [2] the *e-p* wave-function is given by:

$$\psi_p^{e-p}(x_e, x_p) = \exp(ip \cdot x_e) + \sum_{\tilde{p} > p_F} \zeta(p, \tilde{p}) \exp(i\tilde{p} \cdot x_e) \exp[i(p - \tilde{p}) \cdot x_p],$$

where function $\zeta(p, \tilde{p})$ describes a perturbation (due to the *e-p* interaction) of the free-electron state *p*. Due to the Pauli principle, in the case of the electron gas where all states inside the FS are fully occupied, perturbated states can be described only by $\tilde{p} > p_F$. Resulting enhancement factor is growing with *p*, having maximal values at the FS. So, such Kahana-like enhancement is for the case when all states inside the FS are fully occupied (the probability of scattering is the highest for electrons at the FS). However, in real solids, due to the lattice potential, for each occupied band, there is always a leading term of density (where the occupation number is lower than 1) and the Umklapp components. So, one could expect the following: the higher the lattice effects are the weaker is the Kahana-like *p* dependence of the enhancement. This fact is an inherent feature of the BML theory where the *e-p* interaction matrix is based on electron Bloch eigenstates. Therefore, in this theory, the influence of the crystal lattice on *e-e* and *e-p* scattering processes is more realistically described than in other theoretical approaches.

Moreover, it is not true that the dynamic parts of the direct *e-p* and *e-e* interactions cancel each other and in the positron annihilation experiment one should observe only the static part of the *e-p* interaction. *e-e* correlation are not cancelled, they are strong (as in the Compton scattering experiment [26]) and they cannot be described by the isotropic LP correction [25].

**Acknowledgements.** We are very grateful to the Polish State Committee for Scientific Research (Grant 2 P03B 012 25) for the financial support.